# Exploring Content-Based and Meta-Data Analysis for Detecting Fake News Infodemic: A case study on COVID-19

Oluwaseun Ajao, Ashish Garg and Márjory Da Costa-Abreu

*Department of Computing*
*Sheffield Hallam University*
s.ajao@shu.ac.uk; m.da-costa-abreu@shu.ac.uk

*Abstract*—The coronavirus pandemic (COVID-19) is probably the most disruptive global health disaster in recent history. It negatively impacted the whole world and virtually brought the global economy to a standstill. However, as the virus was spreading, infecting people and claiming thousands of lives so was the spread and propagation of fake news, misinformation and disinformation about the event. These included the spread of unconfirmed health advice and remedies on social media. In this paper, false information about the pandemic is identified using a content-based approach and metadata curated from messages posted to online social networks. A content-based approach combined with metadata as well as an initial feature analysis is used and then several supervised learning models are tested for identifying and predicting misleading posts. Our approach shows up to 93% accuracy in the detection of fake news related posts about the COVID-19 pandemic.

## I. INTRODUCTION

On 11th March 2020, the world health organisation (WHO) declared COVID-19 a pandemic. At the time of writing this article, there have been almost 500 million confirmed cases worldwide, and more than 6 million people have now died after they were infected by the disease [1].

An event of this scale has been widely discussed on TV media as well as online social networks due to the global impact [2]. The increased activity witnessed on social media platforms was further driven by 'social distancing' measures being put in place by most affected countries to limit face-to-face contact and aim to prevent further spread of the disease.

Figure 1 shows that social media platforms such as Facebook, YouTube, Twitter and WhatsApp have the highest monthly average users at the end of July 2021 [3], with over 6 billion users combined and the increased usage of these information sharing mediums also brought along with it a challenging surge of false information being circulated amongst users.

There has been a wide range of 'fake news' related with the coronavirus pandemic, ranging from 'miracle cures', blame-game and 5G implications. Moreover, the impact in each country, even though the disease is the same, the population behaviour related to the 'fake news' has been very different.

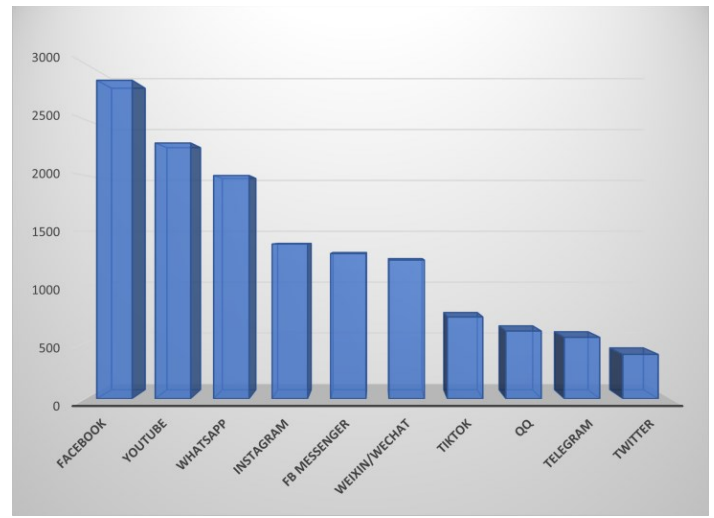

Fig. 1. Social media monthly average users (millions) [3]

The platform Twitter has played an important role on this dissemination [4]. Twitter updates its rules and enforcement policy in an attempt to curtail the spread of misleading health messages messages by increasing control and making additional checks of tweets that gave potentially unfounded, harmful health advice or promoted online hate amongst users regarding the pandemic [5].

However, the dissemination of fake news regarding COVID-19 has never been higher and there is an urgent need to explore the possibilities on how to build a reliable solution. Thus, this paper aims to explore the possible content-based feature extraction as well as the metadata and which machine learning models of supervised learning can indicate which families of solutions might have a bigger impact on this task.

The rest of this paper is structured as follows: Section II explains the main differences on the terminology and how we will proceed in this investigation. Section III presents the most common features and classifiers that are currently being used for COVID-19 related fake news detection. Section IV presents our approach to this work regarding data source and classification. And, finally, Section V presents our conclusions.

## II. UNDERSTANDING THE TERMINOLOGIES: MISINFORMATION, MAL-INFORMATION AND DISINFORMATION

The propagation of 'fake news' is not a new behaviour. There are related work that show its impact on other situations, especially when linked with the popularity of social media and instant messaging applications [6].

In understanding the propagation of false information, three definitions are established [7]:

- **Mis-information** represents messages that contained unconfirmed or untrue information circulated without a deceitful intent but not fact-checked or verified by credible sources before it is disseminated to other users.
- **Dis-information** a type of false information circulated with a deliberate intention to deceive or mislead the recipient of such messages.
- **Mal-information** is false information that is intended to make an inaccurate claim sound believable. It is often associated with testimonials given by alleged specialists within a domain. The motive is first to cause harm to the target and possibly profit or gain for the creator(s) of this message

One new word recently derived from the outbreak is **Infodemic**. WHO [8] has defined the term:

> "An infodemic is an over-abundance of information – some accurate and some not – that makes it hard for people to find trustworthy sources and reliable guidance when they need it."

The dissemination of information regarding any topic is important and can make public engagement more effective. One of the first studies relating to the impact of fake news in social media dissemination was done by analysing data from Weibo [9] and already indicates some hard truths. However, due to the diversity of public knowledge and beliefs, the types of information being propagated amongst users can be either positive or negative, especially during global situations such as the coronavirus pandemic. [10].

Moreover, there is a strong indication that most people who tend to share 'fake news' do it simply because they fail to evaluate the veracity of the information [11].

It is possible to evaluate the quality and authenticity of information in online social networks [12]. It has already been shown that during the COVID-19 outbreak, there has been an increase in the spread of hate speech online including Islamophobia and racism - especially targeted towards people from China - where the virus is believed to have originated [13].

There have been several past studies about automated fact-checking of fake news stories as well as fake news detection using machine learning and artificial intelligence models [14], [15], [16], [17], [18], [19], [20]. As well as COVID misinformation datasets [21], [22], [23].

It should be noted that at the time of writing this paper, we do not aim to distinguish between the various types of fake news i.e. Misinformation, Disinformation and Mal-information. However, the scope of this paper is limited to identifying false social media posts about the COVID-19 pandemic using a combination of the features within the message text and meta-data information.

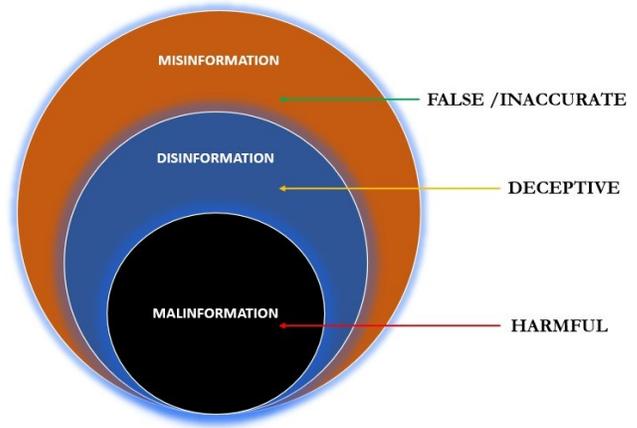

Fig. 2. Malinformation, Misinformation and Disinformation relationships

The main difference between those categories of false posts is synonymous with the intention of the author. While misinformation is always a false post created, shared or circulated with or without an aim of deception, a disinformation post on the other hand would be propagated with the deliberate intention to deceive or mislead the recipient of the message. Hence, disinformation posts are themselves a subset of misinformation posts. While malinformation is also a subset of disinformation as illustrated in Figure 2.

It is important to note that while it is difficult to understand the motives of an author, the context of the message text and any other media associated with it, such as images and video can give an indication with regards to the motive of the author. For the purpose of this study, all three types of false posts (misinformation, disinformation and mal-information) would be regarded as misinformation. They are messages that have been confirmed as false sometimes based on fact-checking websites such as *factcheck.org*, *fullcheck.org*, *politifact.com* and *snopes.com*.

## III. 'FAKE NEWS' AND COVID-19: WHAT THE CURRENT LITERATURE PRESENTS

As mentioned in the previous section, the awareness regarding the spread of 'fake news' using social networks is relatively new, however, there is still a lot to tackle in order to have an efficient and effective mechanism to both identify users and posts that propagate this type of *pandemic*.

At the start of the pandemic, it was not surprising to have the dissemination of 'fake news' regarding COVID-19. Social media has always been a useful measure of public awareness. For example during the outbreak of the H7N9 influenza and MERS-CoV outbreak, there was fake news being shared about the event online. Hence this was expected for the COVID-19 pandemic as well [24]. Also, it is important to note that there is now a widespread impact on knowledge about coronavirus in different countries due to the dissemination of 'fake news' information [25].

In this section, we will focus on presenting the approaches and techniques that have been employed so far for combating 'fake news' dissemination about coronavirus, regardless of it being mis-information, dis-information or mal-information.

### A. Data sources and datasets

There has been significant effort from the research community regarding availability of COVID misinformation to support further studies [26]. Social media datasets used in this study comes from two publicly available and peer-reviewed Twitter datasets by Bandal *et al* [27] and Kouzy *et al* [28].

Despite the fact that the aim of this work is to investigate the use of features as well as machine learning models applied to 'fake news' detection on social media related to coronavirus, some of this fake information can come from medical data falsified clinical results presented as genuine trial results datasets. Some of these have also been made publicly available and peer-reviewed as well, such as [29].

### B. Features

One important aspect of any machine learning-based system is to select and use the appropriate features for model classification. Thus, in this section, we will focus on presenting the current trends related to the features used and the techniques adopted for feature extraction. There are two main approaches for 'fake news' analysis. This could either be with regards to identifying the author - who posted the 'fake news' or in terms of content - identifying if a 'fake news' messages was posted. [30]. Both approaches can be done either independently or combined. However, most studies tend to focus on only one of those two approaches. This work focuses on the content-based approach. As such, the direct identification of authors of these misinformation posts is beyond the scope of the current study and would be considered in future work.

The simplest way to identify relevant characteristics that can be associated with the dissemination of 'fake news' is to analyse the metadata [31]. In our context, metadata can be defined as any information that is not 'available within the message text for semantic interpretation' and it can be automatically calculated from the datasets, such as: number of mentions, number of emojis, number of friends, age of the account, geo-location of the account, geo-location of the post, number of sharing related to that post, number of likes in that post, etc. Such information can be a powerful tool and it can be used in order to narrow any search related to any specific problems.

The most popular way, even though less reliable, is the use of content-based analysis that often happens with the help of natural language processing techniques, such as: lexical semantics tasks and across many parameter settings [32], text analytic–driven approach [33] and graph network characteristics measures [5].

Even though, there are several approaches and techniques that have been used to identify 'fake news', to the best of our knowledge our work is the first to look at incorporating features from both the content and metadata of tweets in predicting misinformation about the pandemic.

Thus, in our work, we explore the use of several different features and analyse their impact on simple machine learning models, ensembles and deep learning models specific to the case of 'fake news' dissemination related with COVID-19 and coronavirus. The main difference between this work and the previous studies relies on the inclusion of the meta-data, the selected machine learning models, and the datasets used.

### C. Techniques used to perform fake news detection

Since 'fake news' on social media has become a potential source of harm in the society, it is also important to understand the machine learning models that can be used for this and how these models can impact on the decision making process for the policy makers as well as the companies that develop those platforms.

As with the feature extraction and selection techniques, it is only natural that the machine learning models for 'fake news' detection in the COVID-19 datasets would be similar to the ones used previously in other hate speech applications. The most popular model for this end that we can already see in the literature is the Long Short-Term Memory (LSTM) network [34]. However, some new modifications for the LSTM can be already seen to be used for text semantic interpretation such as Latent Dirichlet Allocation [35].

When we look at the traditional supervised learning models, the classifier which has been most popular is the SVM. It could either be the multi-class or the single class approach. [36]. Not many other models have been widely used with success so far for COVID-19 classification. Another fusion approach that can also be mentioned is the use of multi-agent systems with game theory variations [37]. Based on what has been presented in the previous sections, we can identify that there is still room for investigation of which machine learning models work better for the identification of 'fake news' related with COVID-19 as well as the identification of the users which are more likely to be sharing such content.

## IV. OUR METHODOLOGY: TWITTER ANALYSIS FOR COVID-19 RELATED FAKE NEWS

Since our work is focusing on analysing the real impact on the features used for 'fake news' identification, we have decided to explore the main techniques that are popular for feature extraction from both content analysis as well as metadata and a few well know classifiers ranging from single models, ensembles and deep learning.

Most data sets of fake news available are from Twitter [38] and are related with user exploitation and hate crime, endangerment of public health, as well as public mistrust. An Oxford university research led by Brennen et al. [39] found that 59% of coronavirus pandemic related fake news remains on Twitter with no warning label. That reinforces that there is a scope of bringing an effective mechanism that detects the fake news automatically inside the tweets.

Thus, we have used a diverse COVID-19 healthcare-related fake news dataset called CoAID (Covid-19 heAlthcare mIsinformation Dataset) [23]. Tweets such as "Only older adults and young people are at risk", "The virus will die off when temperatures rise in the spring" and "Antibiotics kill coronavirus" are few of the many COVID-19 fake news messages that CoAID contains.

This data-set has the 183, 564 tweets where the titles of news articles have used as a search query with start and end dates of the tweets. Table I shows the fake versus real bifurcation in the collected tweets. News article is the presentation of the fake (misleading) or true claims as news. Claims is much shorter than news article and contains only one or two sentences [23]. This data-set is already labelled as the fake and real.

| Source | False Claim | Fake News Article | True Claim | True News Article | TOTAL |
|--------|-------------|-------------------|------------|-------------------|-------|
| Tweets | 457 | 9218 | 6342 | 87324 | 103341 |
| Replies | 623 | 5721 | 9764 | 64115 | 80223 |
| TOTAL | 1080 | 14939 | 16106 | 151439 | 183564 |

TABLE I
TWEETS DATA DISTRIBUTION

The target imbalance in the dataset was quite evident with 96:4 ratio when evaluated. There are two main methods deal with unbalanced datasets such as under sampling, over sampling [40]. Under sampling cuts the majority class data to ensure the proper balance between the classes, speeding up the process but bound to information loss from abundant class if not properly been used. On the other hand, over sampling can increase the rare class dataset to remove imbalance among the classes by retaining all the information of data-set but many times it increases the likelihood of over-fitting when it replicates the minority class data (random over-sampling) [41].

The method used in our study to balanced the unbalanced dataset is one sided selection (OSS) which is an informed under-sampling type [41] [42]. This method selects the representative subset of the majority class 'Non-Fake tweets' and ultimately combines it with the set of all minority class 'Fake tweets' [43]. The subset ratio selected here was 30:70 of minority and majority classes respectively and the final count of tweets after re-balancing comes to 21, 527 samples. The ration was based on the method where experimented data-set with different ratios [44]

The Twitter policy sets a restriction on tweet content redistribution, due to which the only available tweet data comprised the tweet ids [45]. In order to retrieve the required information from the tweet id and to re-hydrate the remaining valuable tweet content, we have used the Twitter APIs. Our feature extraction model for information finding technique used the content-based feature where information was collected directly from the tweets with linguistic characteristics. Specifically, features used here are topic-based, message based and user based and are aggregates measured from the feature sets; for example, Tweet text, tweet-location, hashtags and tagged users [46]. The same is highlighted in the Table II.

| Feature | Description | Type |
|---------|-------------|------|
| TWEET TEXT | Derived Tweet Text + Tweet Hashtags + User Mention + User Name + User Location | Message based |
| WORD COUNT | Number of words in the tweet | Topic based |
| RETWEET COUNT | Number of times message retweeted | Topic based |
| HASHTAG COUNT | Number of times particular hashtag used | Topic based |
| HASHTAGS | Hash tags used in the tweets | Message based |
| USER MENTION | Users mention in the tweets | Message based |
| USER MENTION COUNT | Number of times specific user mention in the tweets | Topic based |
| TWEET URL | Url used in the tweets | Message based |
| TWEET URL COUNT | Number of times url used in the tweet | Message based |
| ACCOUNT AGE | Age of the account | User based |
| USER LOCATION | Location from where tweet has been initiated | User based |

TABLE II
TWEETS FEATURES

The feature extraction results have given us the dataset for the tokenisation and vectorisation which can be seen in Table III. Major features have been divided into two categories and used the label encoding technique [47] based on the patterns. For example, word count below 10 is represented by value 0 and value 1 shows the count more than 10 words. Target variable 'Label' has two classes, namely fake (1) and non-fake (0). As a pre-processing step, these features were cleaned first, and then combined with all the text features inside a single Natural Language Processing feature vector with numerical data in another feature vector which is later combined. The targets are the binary labels of the dataset (fake and non-fake) [48].

Some underlying studies [49]–[51] have used different supervised machine learning algorithms for tweet classification. Thus, in evaluating the performance of COVID-19 related fake news detection on the set of determined features, we have selected for our experiments SVM, Random Forest, Logistic Regression, MNB, RNN and CNN. These different machine learning algorithms were applied on the cleaned data which was obtained after the data pre-processing step in the workflow (Figure 4). First step is to ingest the tweets with the target variables (1 for Fake and 0 for Non-Fake). Data cleansing step involved handling missing data and duplicate data. Feature engineering was identified as a critical step where features were checked against target variable if it showed any pattern in a distribution plot or not. By visualizing it made easy to select the features and considered it for the classifier. For example, the word count distribution from (Figure 3) provided an indication that mostly the word count below 10 belonged to the fake tweets and above it were majorly non-fake tweets and hence transformed this numerical data to binary form. The following steps were to perform natural language processing (cleaning of the tweets), data normalization (to convert the numerical features into a standardised format), vectorization (conversion of the character variables into word embeddings of vector matrices). Afterwards, model training, testing and evaluation processes would performed (using the machine learning and AI classifiers). Finally model prediction is done.

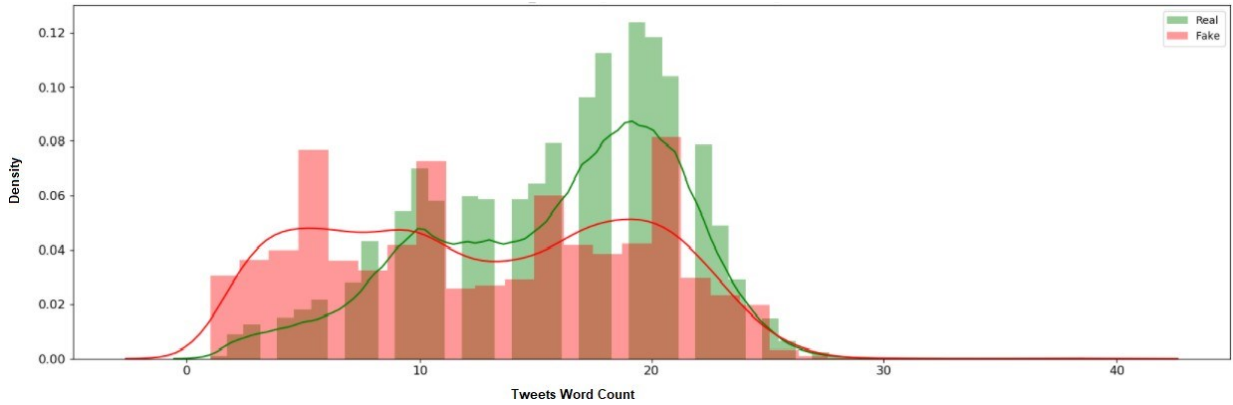

Fig. 3. Word Count Distribution of Training Tweets

| Text | Is Usr Vrfd | Word Cnt | Tweet URL Cnt | Hash tag Cnt | Usr Mention Cnt | Accnt Age | label |
|---|---|---|---|---|---|---|---|
| Only older adults and young people are at r... | 0 | 0 | 1 | 1 | 0 | 1 | 1 |
| SOME MYTHS ABOUT COVID-19. If you can hold... | 1 | 1 | 0 | 1 | 0 | 0 | 1 |

TABLE III
Some Feature Extraction Examples

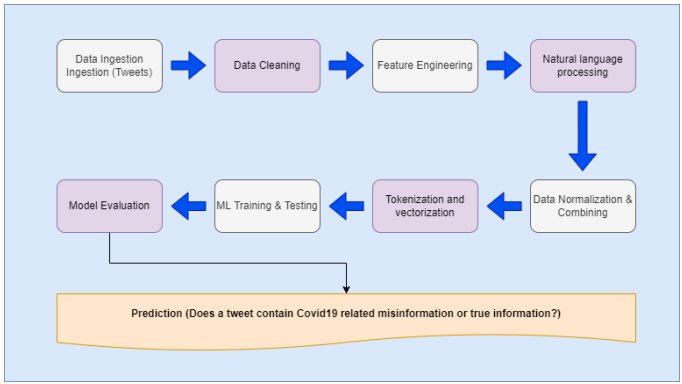

Fig. 4. Data Preprocessing Steps

A 5-fold cross-validation technique has been found to give more reliable results in classification tasks [52]. As a result, this was adopted in the current piece of work. Term Frequency-Inverted Document Frequency (TF-IDF) is picked as a content-based feature extraction technique because as it is useful in the experimental evaluation of the classification problem [53]. During the implementation of the current work, it has been identified that when we select 5,000 tokens across the tweets, we achieved better accuracy results compared to 500 and 1,000. The preliminary results obtained are not shown here due to space constraints. Hence we have selected the token size of 5,000 which has been indicated as the higher TF(Term Frequency) and therefore most significant token in the entire corpus.

Figure 5 shows the results of different machine learning algorithms used as a part of this study. The traditional machine learning model which produced the best evaluation result is the Support Vector machines (SVMs) with accuracy of 93% and precision of 92%. Amongst deep learning models, Recurrent Neural Network (RNN) gave the best result 86% of accuracy. Although SVM has provided the best in terms of result quality. However, it was also identified as the slower classifier model

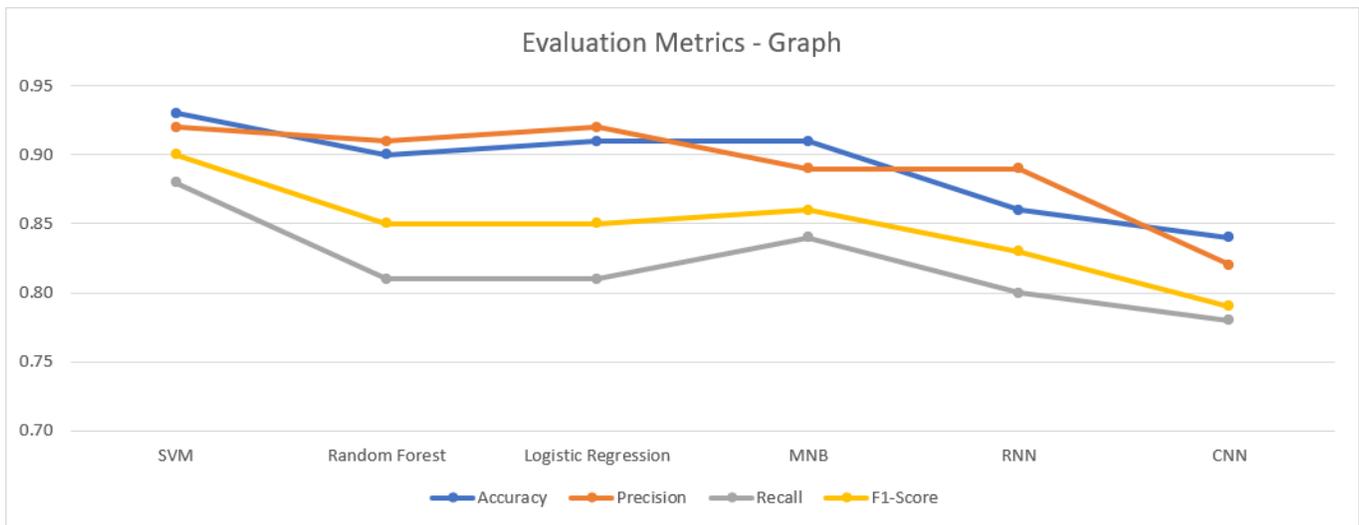

Fig. 5. Model performance comparison

|  | Accuracy | Precision | Recall | F1-Score |
|---|---|---|---|---|
| SVM | 0.93 | 0.92 | 0.88 | 0.90 |
| Random Forest | 0.90 | 0.91 | 0.81 | 0.85 |
| Logistic Regression | 0.91 | 0.92 | 0.81 | 0.85 |
| MNB | 0.91 | 0.89 | 0.84 | 0.86 |
| RNN | 0.86 | 0.89 | 0.80 | 0.83 |
| CNN | 0.84 | 0.82 | 0.78 | 0.79 |

in terms of the processing time.

There is a clear indication that the SVM is the most suitable machine learning algorithm for COVID-19 related fake news detection due to its average f-measure and accuracy result. Various parameters of SVM have been considered for optimising the performance of this algorithm. Parameter optimisation or fine tuning is one of the key activities of the experiment due to which it become possible to achieve the best possible result on the test dataset. [54]. Several parameters of SVM were examined in an attempt to optimise the performance of it. However, the best outcome was achieved with the default parameter settings only.

The experimental results for parameter settings were characterised by thirty measures (5-fold cross-validation gave the best hyper-parameter fit for the accuracy). These results were compared with a simple one-sample t-test (95% confidence interval) statistical method. The null hypothesis which says the average mean accuracy for the SVM is 93% is accepted after the test.

These results suggest that SVM and RNN are the best-performing classifiers among traditional machine learning and deep learning algorithms respectively. They had given the best average f-measure and accuracy score for the COVID-19 related fake news detection on the selected dataset. Also, it is observed that the deep learning techniques also gave some good results albeit not the best.

## V. CONCLUSIONS AND FUTURE WORK

The findings from our study has shown that we can achieve reasonable prediction of more than 90% for both accuracy and precision of the COVID-19 fake news messages from both the content of the tweets and engineering features from the metadata. We also show that using a combined approach of content-based analysis and metadata information, it gives a more suitable feature set which consequently enhances the classification task.

We have looked at binary classification of fake news posts about the pandemic. Future work could look at the detection of multi-class classification based on the different fake news types which we have defined i.e. misinformation, disinformation and mal-information.

Also, direct identification of the authors and origin of fake news posts about health messages such as COVID-19 could be helpful in preventing healthcare malpractices, fraud and support law enforcement.